\pdfoutput=1
\documentclass[sigconf]{acmart}

\usepackage{tugraz_defaults}
\usepackage{enumitem}
\newcommand{\parhead}[1]{\vspace{.5\baselineskip}\noindent\textbf{#1.}\ }

\newcommand{\ValFirstBitflip}{\SI{300}{\ms}}
\newcommand{\ValBitflipRateMax}{\SIx{10000}}

\settoggle{showOutlineNotes}{true}

\begin{document}

\newcommand{\AttackName}{Nethammer\xspace}
\newcommand{\RowhammerSquareHammer}{\tikzsetnextfilename{rowhammer_square_hammer}\begin{tikzpicture}[]
\draw[preaction={fill, cyan}, pattern=north west lines, pattern color=black] (0, 0)  rectangle ++(0.15,0.15);
\end{tikzpicture}%
}
\newcommand{\RowhammerSquareBitflip}{\tikzsetnextfilename{rowhammer_square_bitflip}\begin{tikzpicture}[]
\draw[preaction={fill, red}, pattern=dots, pattern color=black] (0, 0)  rectangle ++(0.15,0.15);
\end{tikzpicture}%
}

\title{Nethammer: \\ Inducing Rowhammer Faults through Network Requests}
\renewcommand{\shorttitle}{Nethammer}

\author{Moritz Lipp}
\affiliation{%
  \institution{Graz University of Technology}
}

\author{Misiker Tadesse Aga}
\affiliation{%
  \institution{University of Michigan}
}

\author{Michael Schwarz}
\affiliation{%
  \institution{Graz University of Technology}
}

\author{Daniel Gruss}
\affiliation{%
  \institution{Graz University of Technology}
}

\author{Clémentine Maurice}
\affiliation{%
  \institution{Univ Rennes, CNRS, IRISA}
}

\author{Lukas Raab}
\affiliation{%
  \institution{Graz University of Technology}
}

\author{Lukas Lamster}
\affiliation{%
  \institution{Graz University of Technology}
}

\renewcommand{\shortauthors}{M. Lipp~\etal}

\acmDOI{XX.XXX/XXX_X}

\acmISBN{XXX-XXXX-XX-XXX/XX/XX}

\acmConference[CCS'18]{ACM Conference on Computer and Communications
  Security}{October 2018}{Toronto, Canada}
\acmYear{2018}
\copyrightyear{2018}
\acmPrice{15.00}


\begin{abstract}

A fundamental assumption in software security is that memory contents do not change unless there is a legitimate deliberate modification.
Classical fault attacks show that this assumption does not hold if the attacker has physical access. 
Rowhammer attacks showed that local code execution is already sufficient to break this assumption.
Rowhammer exploits parasitic effects in DRAM to modify the content of a memory cell without accessing it.
Instead, other memory locations are accessed at a high frequency.
All Rowhammer attacks so far were local attacks, running either in a scripted language or native code.

In this paper, we present \AttackName.
\AttackName is the first truly remote Rowhammer attack, without a single attacker-controlled line of code on the targeted system.
Systems that use uncached memory or flush instructions while handling network requests, \eg for interaction with the network device, can be attacked using \AttackName.
Other systems can still be attacked if they are protected with quality-of-service techniques like Intel CAT.
We demonstrate that the frequency of the cache misses is in all three cases high enough to induce bit flips.
We evaluated different bit flip scenarios.
Depending on the location, the bit flip compromises either the security and integrity of the system and the data of its users, or it can leave persistent damage on the system, \ie persistent denial of service.

We investigated \AttackName on personal computers, servers, and mobile phones.
\AttackName is a security landslide, making the formerly local attack a remote attack.
With this work we invalidate all defenses and mitigation strategies against Rowhammer build upon the assumption of a local attacker.
Consequently, this paradigm shift impacts the security of millions of devices where the attacker is not able to execute attacker-controlled code.
\AttackName requires threat models to be re-evaluated for most network-connected systems.
We discuss state-of-the-art countermeasures and show that most of them have no effect on our attack, including the target-row-refresh (TRR) countermeasure of modern hardware.
\end{abstract}

\settopmatter{printacmref=false}

%
%

\maketitle

\textbf{Disclaimer}: This work on Rowhammer attacks over the network was conducted independently and unaware of other research groups working on truly remote Rowhammer attacks.
Experiments and observations presented in this paper, predate the publication of the Throwhammer attack by~\citeA{Tatar2018}.
We will thoroughly study the differences between both papers and compare the advantages and disadvantages in a future version of this paper.

\section{Introduction}\label{sec:introduction}

Hardware-fault attacks have been considered a security threat since at least 1997~\cite{Boneh1997,Biham1997}.
In such attacks, the attacker intentionally brings devices into physical conditions which are outside their specification for a short time.
For instance, this can be achieved by temporarily using incorrect supply voltages, exposing them to high or low temperature, exposing them to radiation, or by dismantling the chip and shooting at it with lasers.
Fault attacks typically require physical access to the device.
However, if software can bring the device to the border or outside of the specified operational conditions, software-induced hardware faults are possible~\cite{Tang2017,Kim2014}.

The most prominent hardware fault which can be induced by software is the Rowhammer bug, caused by a hardware reliability issue of DRAM.
An attacker can exploit this bug by repeatedly accessing (\emph{hammering}) DRAM cells at a high frequency, causing unauthorized changes in physically adjacent memory locations.
Since its initial discovery as a security issue~\cite{Kim2014}, Rowhammer's ability to defy abstraction barriers between different security domains has been improved gradually to develop more powerful attacks on various systems.
Examples of previous attacks include privilege escalation, from native environments~\cite{Seaborn2015BH,Gruss2018Rowhammer}, from within a browser's sandbox~\cite{Gruss2016Row,Bosman2016,Frigo2018rowhammer}, and from within virtual machines running on third-party compute clouds~\cite{Xiao2016}, mounting fault attacks on cryptographic primitives~\cite{Razavi2016,Bhattacharya2016}, and obtaining root privileges on mobile phones~\cite{Vanderveen2016}.

Most Rowhammer attacks assume that two DRAM rows must be hammered to induce bit flips.
The reason is that they assume that an ``open-page'' memory controller policy is used, \ie a DRAM row is kept open until a different row is accessed.
However, modern CPUs employ more sophisticated memory controller policies that preemptively close rows~\cite{Gruss2018Rowhammer}.
Based on this observation, \citeA{Gruss2018Rowhammer} described a technique called \textit{one-location} hammering.

In 2016, Intel introduced Cache Allocation Technology (CAT) to address quality of service in multi-core server platforms~\cite{CacheQoS2016}.
Intel CAT allows restricting cache allocation of cores to a subset of cache ways of the last-level cache, with the aim of optimizing workloads in shared environments, \eg protecting virtual machines against performance degradation due to cache thrashing of a co-located virtual machine.
However, with a lower number of cache ways available to the process, the probability to evict an address by accessing other addresses increases significantly.
\citeA{Aga2017} showed that this facilitates eviction-based Rowhammer attacks.

All previously known Rowhammer attacks required some form of local code execution, \eg JavaScript~\cite{Gruss2016Row,Bosman2016,Frigo2018rowhammer} or native code~\cite{Kim2014,Seaborn2015BH,Xiao2016,Razavi2016,Bhattacharya2016,Vanderveen2016,Aweke2016,Qiao2016,Aga2017,Mutlu2017rowhammer,Gruss2018Rowhammer}.
Moreover, all works on Rowhammer defenses assume that some form of local code execution is required~\cite{Kim2014,Irazoqui2016mascat,Herath2015,Payer2016,Gruss2016Flush,Corbet2016,Chiappetta2015,Zhang2016CloudRadar,Brasser2017catt,WindowsServerDeduplication,RedHatKSM,Bosman2016,Aweke2016,Kim2015,Ghasempour2015}.
In particular, we found that none of these works even mentions the theoretical possibility of truly non-local Rowhammer attacks.
Consequently, it was a widely accepted assumption that remote Rowhammer attacks are not possible.
More specifically, devices where an attacker could not obtain local code execution were so far considered to be safe.
Yet, the following questions arise: 

{\centering{\emph{
Are remote Rowhammer attacks possible?
More specifically, is it possible for an attacker to induce bit flips and exploit them, without any local code execution on the system?
}
}}

In this paper, we answer these questions and confirm that truly remote Rowhammer attacks are possible.
We present \AttackName, the first Rowhammer attack that does not require local code execution.
\AttackName requires only a fast network connection between the attacker and victim.
It sends a crafted stream of size-optimized packets to the victim which causes a high number of memory accesses to the same set of memory locations.
If the network driver or other parts of the network stack use uncached memory or flush instructions, \eg for interaction with the network device, an attacker can induce bit flips.
Furthermore, if Intel CAT is activated, \eg as an anti-DoS mechanism, memory accesses lead to fast cache eviction and thus frequent DRAM accesses.
This enables attacks even if there are no accesses to uncached memory or flush instructions while handling the network packet.   
Thus, the attacker implicitly hammers the DRAM through the code executed for processing the network packets.
While an attacker cannot control the addresses of the bit flips, we demonstrate how an attacker can still exploit them.

\AttackName has several building blocks that we systematically developed.
First, we measure whether handling network packets could at least, in theory, induce bit flips, and the influence of real-world memory-controller page policies.
For this purpose, we present a new algorithm to observe and classify the memory-controller page policy.
Second, based on these insights, we demonstrate that one-location hammering~\cite{Gruss2018Rowhammer} does not require a closed-page policy, but instead, adaptive policies may also allow one-location hammering.
Third, we investigate memory operations that occur while handling network requests.
Fourth, we show that the time windows we observe between memory accesses from subsequent network requests enable Rowhammer attacks.

As previous work on Rowhammer showed, once a bit flips in a system, its security can be subverted.
We present different attacks exploiting bit flips on victim machines to compromise various services, in particular, version-control systems, DNS servers and OCSP servers.
In all cases, the triggered bit flips may induce persistent denial-of-service attacks by corrupting the persistent state, \eg the file system on the remote machine.
In our experiments, we observed bit flips using \AttackName already after \ValFirstBitflip~of running the attack and up to \ValBitflipRateMax~bit flips per hour.
\AttackName represents a significant paradigm shift, from local to remote attacks.
Previous fault attacks required physical access or local code execution in the case of Rowhammer.
Making Rowhammer possible over the network requires re-evaluating the threat model of virtually every network-connected system.
We discuss state-of-the-art countermeasures and show that most of them do not affect our attack, including the target-row-refresh (TRR) countermeasure in hardware.
Furthermore, we evaluate the performance of different other proposed Rowhammer countermeasures against \AttackName.
\AttackName is difficult to detect on systems where high network traffic is commonplace.
Finally, we discuss how attacks like \AttackName can be mitigated.

\parhead{Contributions} The contributions of this work are:
\begin{itemize}[nolistsep, align=left, leftmargin=0pt, labelwidth=0pt, itemindent=5pt]
\item We present \AttackName, the first truly remote Rowhammer attack, with not even a single line of attacker-controlled code running on the target device.
\item We demonstrate \AttackName on devices that either use uncached memory or \texttt{clflush} while handling network packets.
\item We demonstrate that even without uncached memory and \texttt{clflush}, attacks on cloud systems can still be practical.
\item We illustrate how our attack invalidates assumptions from previous works, marking a paradigm shift, and requiring re-evaluation of the threat models of most network-connected systems.
\item We show that many previously proposed defenses, \eg TRR, do not work against our new attack.
\end{itemize}

\parhead{Outline}
The remainder of the paper is structured as follows.
In~\cref{sec:background}, we provide background information.
In~\cref{sec:attack}, we overview the \AttackName attack.
In~\cref{sec:bridging_the_gap}, we describe the building blocks and obtain insights we need for \AttackName.
In~\cref{sec:exploiting-bit-flips}, we demonstrate how bit flips induced over the network can be exploited.
In~\cref{sec:evaluation}, we evaluate the performance of \AttackName in different scenarios on several different systems.
In~\cref{sec:countermeasures}, we discuss and propose countermeasures.
In~\cref{sec:discussion}, we discuss limitations of \AttackName.
We conclude our work in~\cref{sec:conclusion}.

\section{Background}\label{sec:background}

In this section, we provide the necessary background information on DRAM, memory controller policies, and the Rowhammer attack.
Furthermore, we discuss caches and cache eviction as well as the Intel CAT technology.

\subsection{DRAM and Memory Controller Policies}\label{sec:background:dram}
\parhead{DRAM Organization}
Modern computers use DRAM as the main memory.
To maximize data transfer rates, DRAM is organized for a high degree of parallelism, in a hierarchy of channels, DIMMs, ranks, bank groups, and banks.
Most processors today support dual-channel or quad-channel configurations.
The DIMMs are assigned to one of the channels.
Each DIMM has one or more ranks, \eg the two sides of the DIMM may form two ranks.
Every rank is further subdivided into so-called \textit{banks}, with each bank spanning over multiple chips. 
The number of banks in a rank is standardized~\cite{jedec}, \eg 8 banks on DDR3 and 16 banks on DDR4.
Each bank is an array of \textit{cells}, organized in \textit{rows} and \textit{columns}, storing the actual memory content.
The row size, \ie the amount of data that can be stored in all cells of one row, is defined to be \SI{8}{\kilo B}~\cite{jedec}.
Each cell is made out of a capacitor and an access transistor.
The charge of the capacitor represents the binary data value of the cell.
Each cell in the grid is connected to the neighboring cells with a wire forming horizontal and vertical bit lines.

When accessing a physical address, the memory controller translates the physical address to channel, DIMM, rank, bank group, bank, row, and column addresses.
While AMD publicly documents these addressing functions~\cite{AMDbioskernelguide}, Intel and ARM do not.
\citeA{Pessl2016} reverse-engineered these addressing functions using an automated technique for several Intel and ARM processors.

As DRAM cells lose their charge over time, they must be refreshed periodically.
The maximum time interval between refreshes is defined through the \textit{row refresh rate}, standardized by the JEDEC group for the different DRAM technologies~\cite{jedec}.
Typically, the refresh interval is \SI{64}{\ms} but can vary depending on the device, on-the-fly adjustments due to the current temperature, or other external influences.
With a \SI{64}{\ms} refresh interval, the memory controller issues the refresh command every \SI{7.8}{\us} for each bank.

\parhead{Memory Controller Policies}
Each bank has a \textit{row buffer}, acting as a directly-mapped cache for the rows.
To read data, the data is moved from the cells of a row to the row buffer before it is sent to the processor.
Similarly, write accesses go to the row buffer instead of directly to the row.
By raising the word line of a row, all access transistors in that row are activated to connect all capacitors to their respective bit line.
This transfers the charge representing the data from the row to the row buffer.
If the requested data from this bank is already stored in the row buffer, the data can be transmitted to the processor immediately, resulting in a fast access time (a \textit{row hit}).
However, if the requested data is not in the row buffer, a so-called \textit{row conflict} occurs, and the bit lines must be \textit{pre-charged} before the data can be read from the new target row (row-activate).

Consequently, there are three different cases leading to distinct access times:
row hits are the fastest, an access to a row in a pre-charged bank is a few nanoseconds slower, row conflicts are significantly slower (\ie several nanoseconds).
Hence, the memory controller can optimize the memory performance by deciding when to close a row preemptively and pre-charge the bank.
Typically, memory controllers employ one of the three following page policies:

\begin{enumerate}[nolistsep, align=left, leftmargin=0pt, labelwidth=0pt, itemindent=5pt]
  \item \textbf{Closed-page policy}: the page is immediately closed after every read or write request, and the bank is pre-charged and, thus, ready to open a new row (page-empty).
    If subsequent accesses are likely to be from other rows, a closed-page policy can achieve a better average system performance.
  \item \textbf{Fixed open-page policy}: the page is left open for a fixed amount of time after a read or write request.
    If temporal locality is given, subsequent accesses are served with a low latency.
    This policy is also beneficial for power consumption and bank utilization~\cite{Kaseridis2011}.
  \item \textbf{Adaptive open-page policy}: the adaptive open-page policy by Intel~\cite{Dodd2003} is similar to the fixed open-page policy but dynamically adjusts the page timeout interval.
    Each row buffer has a timeout counter and a timeout register.
    A row remains open until the timeout counter reaches the value of the timeout register.
    As the initial timeout register value might not be the most efficient, an additional mistake counter is introduced to update the timeout register dynamically~\cite{Ghasempour2015}.
    If a row conflict occurs, the memory controller kept the row open for too long and hence, the mistake counter is decremented.
    Whenever a page-empty access could have been a hit as the requested row is the same as the last accessed one, the mistake counter is incremented.
    Periodically, the value of the mistake counter is checked to decide if a less or more aggressive close-page policy should be used.
    If the mistake counter is higher than a certain threshold, the timeout register is incremented to keep the row open for a longer period of time, and conversely, if the mistake counter is lower than a certain threshold, the timeout register is decremented to close the row earlier.
\end{enumerate}

As modern processors have many cores running independently as well as deploy large caches and complex algorithms for spatial and temporal prefetching, the probability that subsequent memory accesses go to the same row decreases.
\citeA{Awasthi2011} proposed an access-based page policy that assumes a row receives the same number of accesses as the last time it was activated.
\citeA{Shen2014} proposed a policy taking past memory accesses into account to decide whether to close a row preemptively.
Intel suggested predicting how long a row should be kept open~\cite{Kareenahalli2003,Chee2003}.
Consequently, more complex memory controller policies have been proposed and are implemented in modern processors~\cite{Kaseridis2011, Ghasempour2015}.
Besides these memory controller policies, the memory controller can also reorder and combine memory accesses~\cite{Rotithor2006}.

\subsection{Rowhammer}\label{sec:background:rowhammer}

\begin{figureA}[t]{rowhammer_strategies}[Different hammering strategies: blue rectangles (\protect\tikzsetnextfilename{rowhammer_square_hammer}\begin{tikzpicture}[]
\draw[preaction={fill, cyan}, pattern=north west lines, pattern color=black] (0, 0)  rectangle ++(0.15,0.15);
\end{tikzpicture}%
) represent the hammered location, while red rectangles (\protect\tikzsetnextfilename{rowhammer_square_bitflip}\begin{tikzpicture}[]
\draw[preaction={fill, red}, pattern=dots, pattern color=black] (0, 0)  rectangle ++(0.15,0.15);
\end{tikzpicture}%
) represent the most likely location for bit flips to occur.]
\begin{center}
\begin{subfigure}[]{0.3\hsize}
  \centering
  \resizebox {\hsize} {!} {
    \tikzsetnextfilename{rowhammer_single_sided}
    \begin{tikzpicture}[]
  \draw[preaction={fill, cyan}, pattern=north west lines, pattern color=black] (1.0, 2.0)  rectangle ++(3.5,0.5);
  \draw[preaction={fill, red}, pattern=dots, pattern color=black] (1.0, 0.5)  rectangle ++(3.5,0.5);
  \draw[preaction={fill, cyan}, pattern=north west lines, pattern color=black] (1.0, 0.0)  rectangle ++(3.5,0.5);
  \draw[preaction={fill, red}, pattern=dots, pattern color=black] (1.0, -0.5)  rectangle ++(3.5,0.5);
  \draw[preaction={fill, cyan}, pattern=north west lines, pattern color=black] (1.0, -2.0)  rectangle ++(3.5,0.5);

  \draw[] ++(0,-1.5) -- ++(5,0);
  \draw[] ++(0,-1.0) -- ++(5,0);
  \draw[] ++(0,-0.5) -- ++(5,0);
  \draw[] ++(0,0.0) -- ++(5,0);
  \draw[] ++(0,0.5) -- ++(5,0);
  \draw[] ++(0,1.0) -- ++(5,0);
  \draw[] ++(0,1.5) -- ++(5,0);
  \draw[] ++(0,2.0) -- ++(5,0);

  \draw[] ++(1.0,-2.0) -- ++(0,4.5);
  \draw[] ++(1.5,-2.0) -- ++(0,4.5);
  \draw[] ++(2.0,-2.0) -- ++(0,4.5);
  \draw[] ++(2.5,-2.0) -- ++(0,4.5);
  \draw[] ++(3.0,-2.0) -- ++(0,4.5);
  \draw[] ++(3.5,-2.0) -- ++(0,4.5);
  \draw[] ++(4.0,-2.0) -- ++(0,4.5);
  \draw[] ++(4.5,-2.0) -- ++(0,4.5);

  \draw (0,-2.0) ++(0,0.25) node[anchor=west] {x$_n$};
  \draw (0,-1.5) ++(0,0.25) node[anchor=west] {\ldots};
  \draw (0,-1.0) ++(0,0.25) node[anchor=west] {x-2};
  \draw (0,-0.5) ++(0,0.25) node[anchor=west] {x-1};
  \draw (0,0.0) ++(0,0.25) node[anchor=west] {x};
  \draw (0,0.5) ++(0,0.25) node[anchor=west] {x+1};
  \draw (0,1.0) ++(0,0.25) node[anchor=west] {x+2};
  \draw (0,1.5) ++(0,0.25) node[anchor=west] {\ldots};
  \draw (0,2.0) ++(0,0.25) node[anchor=west] {x$_m$};
\end{tikzpicture}
  }
  \caption{Single-sided}
\end{subfigure}
\hfill
\begin{subfigure}[]{0.3\hsize}
  \centering
  \resizebox {\hsize} {!} {
    \tikzsetnextfilename{rowhammer_double_sided}
    \begin{tikzpicture}[]
  \draw[preaction={fill, red}, pattern=dots, pattern color=black] (1.0, 1.0)  rectangle ++(3.5,0.5);
  \draw[preaction={fill, cyan}, pattern=north west lines, pattern color=black] (1.0, 0.5)  rectangle ++(3.5,0.5);
  \draw[preaction={fill, red}, pattern=dots, pattern color=black] (1.0, 0.0)  rectangle ++(3.5,0.5);
  \draw[preaction={fill, cyan}, pattern=north west lines, pattern color=black] (1.0, -0.5)  rectangle ++(3.5,0.5);
  \draw[preaction={fill, red}, pattern=dots, pattern color=black] (1.0, -1.0)  rectangle ++(3.5,0.5);

  \draw[] ++(0,-1.5) -- ++(5,0);
  \draw[] ++(0,-1.0) -- ++(5,0);
  \draw[] ++(0,-0.5) -- ++(5,0);
  \draw[] ++(0,0.0) -- ++(5,0);
  \draw[] ++(0,0.5) -- ++(5,0);
  \draw[] ++(0,1.0) -- ++(5,0);
  \draw[] ++(0,1.5) -- ++(5,0);
  \draw[] ++(0,2.0) -- ++(5,0);

  \draw[] ++(1.0,-2.0) -- ++(0,4.5);
  \draw[] ++(1.5,-2.0) -- ++(0,4.5);
  \draw[] ++(2.0,-2.0) -- ++(0,4.5);
  \draw[] ++(2.5,-2.0) -- ++(0,4.5);
  \draw[] ++(3.0,-2.0) -- ++(0,4.5);
  \draw[] ++(3.5,-2.0) -- ++(0,4.5);
  \draw[] ++(4.0,-2.0) -- ++(0,4.5);
  \draw[] ++(4.5,-2.0) -- ++(0,4.5);

  \draw (0,-2.0) ++(0,0.25) node[anchor=west] {\ldots};
  \draw (0,-1.5) ++(0,0.25) node[anchor=west] {x-3};
  \draw (0,-1.0) ++(0,0.25) node[anchor=west] {x-2};
  \draw (0,-0.5) ++(0,0.25) node[anchor=west] {x-1};
  \draw (0,0.0) ++(0,0.25) node[anchor=west] {x};
  \draw (0,0.5) ++(0,0.25) node[anchor=west] {x+1};
  \draw (0,1.0) ++(0,0.25) node[anchor=west] {x+2};
  \draw (0,1.5) ++(0,0.25) node[anchor=west] {x+3};
  \draw (0,2.0) ++(0,0.25) node[anchor=west] {\ldots};
\end{tikzpicture}
  }
  \caption{Double-sided}
\end{subfigure}
\hfill
\begin{subfigure}[]{0.3\hsize}
  \centering
  \resizebox {\hsize} {!} {
    \tikzsetnextfilename{rowhammer_one_location}
    \begin{tikzpicture}[]
  \draw[preaction={fill, red}, pattern=dots, pattern color=black] (1.0, 0.5)  rectangle ++(3.5,0.5);
  \draw[preaction={fill, cyan}, pattern=north west lines, pattern color=black] (1.0, 0.0)  rectangle ++(3.5,0.5);
  \draw[preaction={fill, red}, pattern=dots, pattern color=black] (1.0, -0.5)  rectangle ++(3.5,0.5);

  \draw[] ++(0,-1.5) -- ++(5,0);
  \draw[] ++(0,-1.0) -- ++(5,0);
  \draw[] ++(0,-0.5) -- ++(5,0);
  \draw[] ++(0,0.0) -- ++(5,0);
  \draw[] ++(0,0.5) -- ++(5,0);
  \draw[] ++(0,1.0) -- ++(5,0);
  \draw[] ++(0,1.5) -- ++(5,0);
  \draw[] ++(0,2.0) -- ++(5,0);

  \draw[] ++(1.0,-2.0) -- ++(0,4.5);
  \draw[] ++(1.5,-2.0) -- ++(0,4.5);
  \draw[] ++(2.0,-2.0) -- ++(0,4.5);
  \draw[] ++(2.5,-2.0) -- ++(0,4.5);
  \draw[] ++(3.0,-2.0) -- ++(0,4.5);
  \draw[] ++(3.5,-2.0) -- ++(0,4.5);
  \draw[] ++(4.0,-2.0) -- ++(0,4.5);
  \draw[] ++(4.5,-2.0) -- ++(0,4.5);

  \draw (0,-2.0) ++(0,0.25) node[anchor=west] {\ldots};
  \draw (0,-1.5) ++(0,0.25) node[anchor=west] {x-3};
  \draw (0,-1.0) ++(0,0.25) node[anchor=west] {x-2};
  \draw (0,-0.5) ++(0,0.25) node[anchor=west] {x-1};
  \draw (0,0.0) ++(0,0.25) node[anchor=west] {x};
  \draw (0,0.5) ++(0,0.25) node[anchor=west] {x+1};
  \draw (0,1.0) ++(0,0.25) node[anchor=west] {x+2};
  \draw (0,1.5) ++(0,0.25) node[anchor=west] {x+3};
  \draw (0,2.0) ++(0,0.25) node[anchor=west] {\ldots};
\end{tikzpicture}
  }
  \caption{One-location}
\end{subfigure}
\quad
\end{center}
\end{figureA}

With increasing DRAM cell density, the physical size of DRAM cells and their capacitance decreases.
While this has the advantage of higher storage capacity and lower power consumption, cells may be more susceptible to disturbance errors.
Disturbance errors are interferences between cells that cause memory corruption by unintentionally flipping the bit-value of a DRAM cell~\cite{Mutlu2017rowhammer}.

In 2014, \citeA{Kim2014} demonstrated that such bit flips could be reliably triggered in a DRAM row by accessing memory locations in adjacent DRAM rows in a high frequency, a technique known as \textit{row hammering}~\cite{Huang2012}.
Typically, subsequent memory accesses would be served from the CPU cache.
However, in a Rowhammer attack, the cache is bypassed by either using specific instructions~\cite{Kim2014}, cache eviction~\cite{Gruss2016Row,Aweke2016,Aga2017,Frigo2018rowhammer} or uncached memory~\cite{Qiao2016,Vanderveen2016}.

To reliably induce bit flips, different techniques have been proposed using different memory access patterns as illustrated in~\cref{fig:rowhammer_strategies}.
While the name \textit{single-sided hammering} suggests that only one memory location is accessed, \citeA{Seaborn2015BH} accessed 8 randomly chosen memory locations simultaneously.
\citeA{Seaborn2015BH} focused on a typical DDR3 setup with 32 DRAM banks.
Following the birthday paradox, the probability is quite high that at least 2 out of 8 random memory locations map into the same DRAM bank.
By repeatedly accessing these 8 memory locations, the attacker induces row conflicts at a high frequency.
With single-sided hammering, bit flips most likely occur in some proximity to one of the 8 hammered rows.

With \textit{double-sided hammering}, the attacker chooses three rows, where the two outer rows are hammered.
Bit flips most likely occur in the row between the two rows.
Double-sided hammering requires at least partial knowledge of virtual-to-physical mappings.

Finally, \citeA{Gruss2018Rowhammer} proposed \textit{one-location hammering}, in which the attacker only accesses one single location at a high frequency.
The attacker does not directly induce row conflicts but instead keeps re-opening one row permanently.
As modern processors do not use strict open-page policies anymore, the memory controller preemptively closes rows earlier than necessary, causing row conflicts on the subsequent accesses of the attacker.
Bit flips most likely occur in proximity to the hammered row.

Using these techniques, the Rowhammer bug has been exploited in different scenarios.
\citeA{Bhattacharya2016} exploited untargeted bit flips at random locations to produce faulty RSA signatures, allowing the recovery of the secret keys.
However, as bit flips can be reproduced quite reliably, more deterministic attacks have been mounted.
These attacks include privilege-escalation attacks, sandbox escapes and the compromise of cryptographic algorithms.
They have been mounted from sandboxed environments~\cite{Seaborn2015BH}, from native environments~\cite{Seaborn2015BH,Gruss2018Rowhammer}, from virtual machines in the cloud~\cite{Xiao2016,Razavi2016}, as well as from within a web browser running JavaScript~\cite{Gruss2016Row,Bosman2016}.
Furthermore, attacks from native code~\cite{Vanderveen2016} and JavaScript within the browser sandbox~\cite{Frigo2018rowhammer} have been demonstrated on mobile devices.
To reliably induce a bit flip on a specific page, memory spraying~\cite{Seaborn2015BH,Gruss2016Row,Xiao2016}, grooming~\cite{Vanderveen2016}, and page deduplication~\cite{Bosman2016,Razavi2016} have been used.

To develop countermeasures, a large body of research focused on detecting~\cite{Irazoqui2016mascat,Herath2015,Payer2016,Gruss2016Flush,Corbet2016,Chiappetta2015,Zhang2016CloudRadar}, neutralizing~\cite{Brasser2017catt,Vanderveen2016,Gruss2016Row,Bosman2016,Razavi2016}, or eliminating~\cite{Kim2014,Aweke2016,Corbet2016,Brasser2017catt,Kim2015,Ghasempour2015} Rowhammer attacks in software or hardware.
Furthermore, the LPDDR4 standard~\cite{LPDDR4} specifies two features to mitigate Rowhammer attacks: with Target Row Refresh (TRR) the memory controller refreshes adjacent  rows of a certain row and with Maximum Activation Count (MAC) the number of times a row can be activated before adjacent rows have to be refreshed is specified.
However, in 2018, Gruss~\etal\cite{Gruss2018Rowhammer} showed that an attacker can bypass all software-based countermeasures and gain root privileges by mounting a one-location hammering Rowhammer attack from inside an Intel SGX enclave.

\subsection{Caches and Cache Eviction}\label{sec:background:caches}

Caching is a fundamental concept that is used to reduce the latency of various operations, in particular computations and accesses to slower storage.
Hardware caches keep frequently used data from main memory in smaller but faster memories.

\parhead{Cache Organization}
Modern CPUs have multiple levels of caches, varying in size and latency, where the level-1 (L1) cache is the smallest and fastest, and the L3 or last-level cache is the biggest but slowest cache.
Modern caches are organized in cache sets consisting of a fixed number of cache ways.
The cache set is determined by either the virtual or physical address.
Addresses are called \textit{congruent} if they map to the same cache set.
The cache replacement policy decides which of the cache ways is replaced (evicted) when new data has to be loaded into the cache.

On most Intel CPUs, the last-level cache is inclusive, \ie~data present in L1 or L2 cache must also be present in the last-level cache.
Furthermore, the last-level cache is shared among all cores and divided into so-called cache slices.
The hash function that maps physical addresses to slices is not publicly documented but has been reverse-engineered~\cite{Maurice2015RAID,Yarom2015iacr,Inci2016}.

\parhead{Cache Eviction}
To mount a Rowhammer attack, memory accesses need to be directly served by the main memory.
Thus, an attacker needs to make sure that the data is not stored in the cache.
An attacker can use the unprivileged \texttt{clflush} instruction~\cite{Yarom2014} to invalidate the cache line or use uncached memory if available~\cite{Vanderveen2016}.
On devices where no uncached memory and no unprivileged cache flush instruction is available, an attacker can instead evict a cache line by accessing congruent memory addresses~\cite{Gruss2016Row,Lipp2016,Frigo2018rowhammer}, \ie addresses that map to the same cache set and same cache slice.
Merely accessing a large number of different but congruent addresses in an arbitrary order typically does not lead to a high eviction rate.
\citeA{Gruss2016Row} observed that to evict the victim address, a so-called eviction set of attacker-chosen congruent addresses has to be accessed in a specific pattern.
The eviction set does not contain the victim address, which is consequently evicted from the cache.

\begin{figureA}[t]{intel_cat}[When Intel CAT is disabled in (\protect\subref{fig:intel_cat:disabled}), the cache is shared among the virtual machines. In (\protect\subref{fig:intel_cat:enabled}), Intel CAT is configured with 6 ways for VM1, and 1 way for VM2 and VM3.]
\quad
\begin{subfigure}[]{0.4\hsize}
  \centering
  \resizebox {0.8\hsize} {!} {
    \tikzsetnextfilename{intel_cat_1}
      \begin{tikzpicture}[]
  \tikzstyle{grid} = [draw, step=1]
  \tikzstyle{vm1}=[draw, fill=green]
  \tikzstyle{vm2}=[draw, fill=red]
  \tikzstyle{vm3}=[draw, fill=blue]
  \tikzstyle{vmbox}=[draw,minimum width=2cm,minimum height=1cm,line width=3pt]
  \tikzstyle{waybrace}=[decorate,decoration={brace,amplitude=8pt,raise=6pt},line width=2pt] 
  
  \begin{scope} 
   \draw[vm2] (0,0) rectangle +(1,1);
   \draw[vm1] (1,0) rectangle +(3,1);
   \draw[vm2] (4,0) rectangle +(2,1);
   \draw[vm1] (6,0) rectangle +(1,1);
   \draw[vm3] (7,0) rectangle +(1,1);
   \draw[vm1] (0,1) rectangle +(2,1);
   \draw[vm2] (2,1) rectangle +(2,1);
   \draw[vm3] (4,1) rectangle +(1,1);
   \draw[vm2] (5,1) rectangle +(1,1);
   \draw[vm1] (6,1) rectangle +(1,1);
   \draw[vm2] (7,1) rectangle +(1,1);
   \draw[vm2] (0,2) rectangle +(3,1);
   \draw[vm3] (3,2) rectangle +(1,1);
   \draw[vm1] (4,2) rectangle +(2,1);
   \draw[vm3] (6,2) rectangle +(1,1);
   \draw[vm1] (7,2) rectangle +(1,1);
   \draw[vm3] (0,3) rectangle +(1,1);
   \draw[vm1] (1,3) rectangle +(2,1);
   \draw[vm2] (3,3) rectangle +(3,1);
   \draw[vm1] (6,3) rectangle +(1,1);
   \draw[vm2] (7,3) rectangle +(1,1);
   \draw[grid] (0,0) grid (8,4);
   \draw[draw] (-0.2,-1) rectangle +(8.4,5.2);
   \node[draw=none, text width=7.5cm, align=center] at (4,-0.5) {\Huge Last-Level Cache};
   
   \node[vmbox,draw=green] (vm1) at (1,6.5) {\Huge VM1};
   \node[vmbox,draw=red] (vm2) at (4,6.5) {\Huge VM2};
   \node[vmbox,draw=blue] (vm3) at (7,6.5) {\Huge VM3};
   \draw [waybrace] (0,4) -- (8,4) node {};
   \node (nocatbrace) at (4,4.4) {};
   \path [-latex',style={|-|},thick] (vm1.south) edge (nocatbrace);
   \path [-latex',style={|-|},thick] (vm2.south) edge (nocatbrace);
   \path [-latex',style={|-|},thick] (vm3.south) edge (nocatbrace);
  \end{scope}
  
  \end{tikzpicture}
  }
  \caption{CAT disabled}
  \label{fig:intel_cat:disabled}
\end{subfigure}
\hfill
\begin{subfigure}[]{0.4\hsize}
  \centering
  \resizebox {0.8\hsize} {!} {
    \tikzsetnextfilename{intel_cat_2}
      \begin{tikzpicture}[]
  \tikzstyle{grid} = [draw, step=1]
  \tikzstyle{vm1}=[draw, fill=green]
  \tikzstyle{vm2}=[draw, fill=red]
  \tikzstyle{vm3}=[draw, fill=blue]
  \tikzstyle{vmbox}=[draw,minimum width=2cm,minimum height=1cm,line width=3pt]
  \tikzstyle{waybrace}=[decorate,decoration={brace,amplitude=8pt,raise=6pt},line width=2pt] 
  
  \begin{scope}[shift={(10,0)}] 
   \draw[vm1] (0,0) rectangle +(6,4);
   \draw[vm2] (6,0) rectangle +(1,4);
   \draw[vm3] (7,0) rectangle +(1,4);
   \draw[grid] (0,0) grid (8,4);   
   \draw[draw] (-0.2,-1) rectangle +(8.4,5.2);
   \node[draw=none, text width=7.5cm, align=center] at (4,-0.5) {\Huge Last-Level Cache};
   \node[vmbox,draw=green] (vm1) at (1,6.5) {\Huge VM1};
   \node[vmbox,draw=red] (vm2) at (4,6.5) {\Huge VM2};
   \node[vmbox,draw=blue] (vm3) at (7,6.5) {\Huge VM3};
   
   \draw [waybrace,color=green] (0,4) -- (6,4) node {};
   \node (cat1) at (3,4.4) {};
   \path [-latex',style={|-|},thick] (vm1.south) edge (cat1);
   \draw [waybrace,color=red] (6,4) -- (7,4) node {};
   \node (cat2) at (6.5,4.4) {};
   \path [-latex',style={|-|},thick] (vm2.south) edge (cat2);
   \draw [waybrace,color=blue] (7,4) -- (8,4) node {};
   \node (cat3) at (7.5,4.4) {};
   \path [-latex',style={|-|},thick] (vm3.south) edge (cat3);
   
  \end{scope}
  \end{tikzpicture}
  }
  \caption{CAT enabled}
  \label{fig:intel_cat:enabled}
\end{subfigure}
\quad
\end{figureA}

\parhead{Intel CAT}\label{sec:background:caches:cat}
In 2016, Intel introduced Cache Allocation Technology (CAT)~\cite{Intel_vol3} to address quality of service in multi-core server platforms~\cite{CacheQoS2016,CATPerformance2015}.
Intel CAT allows system software to partition the last-level cache to optimize workloads in shared environments as well as to isolate applications or virtual machines in the cloud.
When a virtual machine in the cloud thrashes the cache and therefore decreases the performance of other co-located machines, the hypervisor can restrict this virtual machine to a subset of the cache to retain the performance of other tenants.
More specifically, Intel CAT allows restricting the number of cache ways available to processes, virtual machines, and containers, as illustrated in~\cref{fig:intel_cat}.
However, \citeA{Aga2017} showed that Intel CAT allows improving eviction-based Rowhammer attacks as it reduces the number of accesses required for cache eviction and consequently reduces the time required to evict an address from the cache.

\section{\AttackName Attack}\label{sec:attack}

All previously published Rowhammer attacks rely on some form of code execution on the targeted device, be it the execution of a native binary~\cite{Seaborn2015BH,Aga2017,Gruss2018Rowhammer,Razavi2016}, an application~\cite{Vanderveen2016} or using a scripted language in the web browser, like JavaScript~\cite{Gruss2016Row,Bosman2016,Frigo2018rowhammer}.
In this section, we present \AttackName, the first Rowhammer attack that does not rely on any attacker-controlled code on the victim machine.

\subsection{Attack Overview}\label{sec:attack:overview}

\AttackName sends a crafted stream of network packets to the target device to mount a one-location or single-sided Rowhammer attack by exploiting quality-of-service technologies deployed on the device.
For each packet received on the target device, a set of addresses is accessed, either in the kernel driver or a user-space application processing the contents.
By repeatedly sending packets, this set of addresses is hammered and, thus, bit flips may be induced.
As frequently-used addresses are served from the cache for performance, the cache must be bypassed such that the access goes directly into the DRAM to cause the row conflicts required for hammering.
This can be achieved in different ways if the code that is executed (in kernel space or user space) when receiving a packet,
\begin{enumerate}[nolistsep, align=left, leftmargin=0pt, labelwidth=0pt, itemindent=5pt]
\item \textit{flushes} (and later on reloads) an address;
\item uses \textit{uncached} memory;
\item \textit{evicts} (and later on reloads) an address.
\end{enumerate}
All three scenarios are plausible.
\textit{Uncached} memory is often used on ARM-based devices for interaction with the hardware, \eg access buffers used by the network controller.
Intel x86 processors have the \texttt{clflush} instruction for the same purpose.
We verified that an attack is practical in both scenarios, as we describe in~\cref{sec:evaluation:results}.

As caches are large, and cache replacement policies try to keep frequently-used data in the cache, it is not trivial to mount an eviction-based attack without executing attacker-controlled code on the device.
However, to address quality of service in multi-core server platforms, Intel introduced CAT (\cf \cref{sec:background:caches:cat}), allowing to control the amount of cache available to applications or virtual machines dynamically as illustrated in~\cref{fig:intel_cat}.
If a virtual machine is thrashing the cache, the hypervisor limits the number of cache ways available to this virtual machine to meet performance guarantees given to other tenants on the same physical machine.
Thus, if an attacker excessively uses the cache, its virtual machine is restricted to a low number of ways, possibly only one, leading to a fast self-eviction of addresses.

\subsection{Attack Setup}\label{sec:attack:setup}

In our attack setup, the attacker has a fast network connection to the victim machine, \eg a gigabit connection.
We assume that the victim machine has DDR2, DDR3, or DDR4 memory that is susceptible to one-location (or single-sided) hammering.

\parhead{Personal Computers}
For our attack on personal computers, tablets, smartphones, or devices with similar hardware configuration, we make no further assumptions.

\parhead{Cloud Systems}
For our attack on cloud systems, we assume that the victim is running a virtual machine on a cloud server providing an interface or API accessible over the network.
Furthermore, to prevent denial-of-service situations due to cache thrashing, we assume that the hypervisor on the cloud server uses Intel CAT to constrain the virtual machine of the victim to a subset of the cache.

Note that there are overlaps between the two attack setups.
A personal computer can be susceptible to the attack we describe for the cloud scenario.
Even more likely a cloud system can be susceptible to the attack we describe for personal computers.

\subsection{Inducing Bit Flips over Network}\label{sec:attack:inducing-bitflips}

To induce bit flips remotely, one requirement is to send as many packets as possible over the network in a short time frame.
As defined in~\cref{sec:attack:setup}, we assume that either uncached memory or \texttt{clflush} is used when receiving a network packet or alternatively, that Intel CAT is active on the victim machine.
Thus, every single packet processed by the network stack actively evicts and reloads data from the cache.
By sending many packets, the corresponding addresses are hammered efficiently.

As an example, UDP packets without content can be used, allowing an overall packet size of \SI{64}{\byte}, which is the minimum packet size for an Ethernet packet.
This allows to send up to \SI{1024000}{packets} per second over a \SI{500}{\mega\bit/\second} connection.
\section{From Regular Memory Accesses to Rowhammer}\label{sec:bridging_the_gap}
Naturally, several challenges need to be solved to induce Rowhammer bit flips over the network.
Fundamentally, we need to investigate memory-controller page policies to determine whether regular memory accesses that occur while handling network packets could at least, in theory, induce bit flips.
Note that these investigations are oblivious to the specific technique to access the DRAM row (\ie eviction, flushing, uncached memory).
Hence, in this section, we do not discuss \texttt{clflush}, uncached memory, or eviction strategies with~\cite{Aga2017} or without Intel CAT~\cite{Gruss2016Row,Lipp2016}.
We defer comparisons of \AttackName with these techniques to \cref{sec:evaluation}.
In this section, we focus on the underlying behavior of the memory controller and what this means for possible attacks.

\citeA{Gruss2018Rowhammer} found that the memory-controller page policy has a significant influence on the way the Rowhammer bug can be triggered.
In particular, they found that one-location hammering works and deduced from this that the memory-controller page policy must be similar to a closed-page policy.
Most previous work on Rowhammer assumed an open-row policy~\cite{Kim2014,Seaborn2015BH,Xiao2016,Razavi2016,Bhattacharya2016,Vanderveen2016,Aweke2016,Qiao2016,Aga2017,Mutlu2017rowhammer}.
In~\cref{sec:evaluation:memory-controller-policy}, we propose a method to determine the memory-controller page policy on real-world systems automatically.
We show that one-location hammering does not necessarily need a closed-page policy, but instead, adaptive policies may allow one-location hammering.

Based on these insights, we demonstrate the first one-location Rowhammer attack on an ARM device in~\cref{sec:bridging_the_gap:arm}, and draw the connection to the attack presented by~\citeA{Aga2017}.
Finally, we investigate whether Rowhammer via network packets is theoretically possible.
Network packets do not arrive with the same speed as the memory accesses in an optimized tight loop.

\subsection{Automated Classification of Memory-Controller Page Policies}\label{sec:evaluation:memory-controller-policy}

\citeA{Gruss2018Rowhammer} stated that a requirement for one-location hammering is a policy similar to a closed-page policy.
To get a more in-depth understanding of the memory-controller page policy used on a specific system, we present an automated method to detect the used policy.
This is a significant step forward for Rowhammer attacks, as it allows to deduce whether specific attack variants may or may not work without an empiric evaluation.
\citeA{Pessl2016} reverse-engineered the undocumented mapping functions of physical memory addresses to DRAM channels, ranks and banks.
These mapping functions allow selecting addresses located in the same bank but in a different row.
If we access these addresses consecutively, we will cause a row conflict in the corresponding bank.
This row conflict induces latency for the second access because the currently active row must be closed (written back), the bank must be pre-charged, and only then the new row can be fetched with an activate command.
This side-channel information can not only be used to build a covert communication channel~\cite{Pessl2016}, but as we show, it can also be used to detect the page policy used by the memory controller.

\parhead{Automated Classification of the Page-policy}
We assume knowledge of processor and DRAM timings.
For the DRAM this means in particular, the \texttt{tRCD} latency (the time to select a column address), and the \texttt{tRP} latency (the time between pre-charge and row activation).
These three timings influence the observed latency as follows:
\begin{enumerate}[nolistsep, align=left, leftmargin=0pt, labelwidth=0pt, itemindent=5pt]
\item we consider the case \textbf{page open / row hit} as the base line;
\item in the case \textbf{page empty / bank pre-charged}, we observe an additional latency of \texttt{tRP} over a row hit;
\item in the case \textbf{page miss / row conflict}, we observe an additional latency of $(\texttt{tRP} + \texttt{tRCD})$ over a row hit.
\end{enumerate}
To compute the actual number of cycles we can expect, we have to divide the DRAM latency value by the DRAM clock rate.
In case of DDR4, we have to additionally divide the latency value by factor two, as DDR4 is double-clocked.
This yields the latency in nanoseconds.
By dividing the nanoseconds by the processor clock speed, we obtain the latency in CPU cycles.
Still, as we cannot obtain absolutely clean measurements due to out-of-order execution, prefetching, and other mechanisms that aim to hide the DRAM latency, the actually observed latency will deviate slightly.

As in our test we cannot simply measure the three different cases, we define an experiment that allows to distinguish the different policies.
In the experiment we use for our automated classification, we select two addresses $A$ and $B$ that map to the same bank but different rows.
Using the \texttt{clflush} instruction, we make sure that $A$ and $B$ are not cached, in order to load those addresses directly from main memory.
We base our method on two observations for open-page policies:
\begin{itemize}[nolistsep, align=left, leftmargin=0pt, labelwidth=0pt, itemindent=5pt]
\item[\textbf{Single}] By loading address $A$ an increasing number of times ($n=1..10\,000$) before measuring the time it takes to load the same address on a subsequent access, we can measure the access time of an address in DRAM if the corresponding row is already active. For an open-page policy the access time should be the same for any $n$.
\item[\textbf{Conflict}] By accessing address $A$ and subsequently measuring the access time to address $B$, we can measure the access time of an address in DRAM in the occurrence of a row conflict.
\end{itemize}

Our classification now works by running the following checks:
\begin{enumerate}[nolistsep, align=left, leftmargin=0pt, labelwidth=0pt, itemindent=5pt]
\item If there is no timing difference between the two cases described above (\textbf{Single} with a large $n$ and \textbf{Conflict}), the system uses a closed-page policy.
The closed-page policy immediately closes the row after every read or write request.
Thus, there is no timing difference between these two cases.
The timing observed corresponds to the row-pre-charged state.
\item Otherwise, if the timing for the \textbf{Single} case is the same, regardless of the value of $n$, but differs from the timing for \textbf{Conflict}, the system uses an open-page policy.
The timing difference corresponds to the row hits and row conflicts.
Following the definition of the open-page policy, the timing for row hits is always the same.
\item Otherwise, the timing for the \textbf{Single} case will have a jump at some $n$ after which the page policy is adapted to cope better with our workload.
Consequently, the timing differences we observe correspond to row hit and row-pre-charged states.
\end{enumerate}

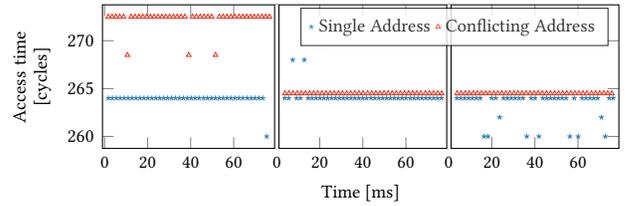
\begin{figureA}[t]{mcp_all}[Measured access times over a period of time for a single address (blue) and an address causing a row conflict (red) for different page policies on the Intel Xeon D-1541: open policy (left), closed policy (middle), adaptive policy (right).]
\resizebox {\hsize} {!} {
    \tikzsetnextfilename{policies}
    \begin{tikzpicture}
\begin{axis}[
style={font=\footnotesize},
width={\hsize},
height=3.5cm,
xlabel={Time [ms]},
ylabel={Access time [cycles]},
ylabel style={text width=2cm,align=center},
xmin=0,
xmax=240,
legend columns=-1,
xtick=      {1,21,41,61,81,101,121,141,161,181,201,221},
xticklabels={0,20,40,60,0,20,40,60,0,20,40,60,80},
legend style={fill=white, fill opacity=0.8, draw opacity=0.8},
]

\addplot+[only marks,mark=star,mark size=1pt,blue, mark repeat=15,mark phase=15] table[x=Time,y=Single, col sep=comma,restrict x to domain=1:78]
  {data/mcp_open_single.csv};
\addplot+[y filter/.code={\pgfmathparse{\pgfmathresult+0.5}}, mark repeat=15,mark phase=15, only marks,mark=triangle,mark size=1pt,red] table[x=Time,y=Single, col sep=comma,restrict x to domain=1:78]{data/mcp_open_conflict.csv};

\addplot+[only marks,mark=star,mark size=1pt,blue, mark repeat=15,mark phase=15] table[x=Time,y=Single, col sep=comma,restrict x to domain=83:158]
  {data/mcp_closed_single.csv};
\addplot+[y filter/.code={\pgfmathparse{\pgfmathresult+0.5}},mark repeat=15,mark phase=15,only marks,mark=triangle,mark size=1pt,red] table[x=Time,y=Single, col sep=comma,restrict x to domain=83:158]{data/mcp_closed_conflict.csv};

\addplot+[only marks,mark=star,mark size=1pt,blue,mark repeat=15,mark phase=15] table[x=Time,y=Single, col sep=comma,restrict x to domain=163:238]
  {data/mcp_adaptive_single.csv};
\addplot+[y filter/.code={\pgfmathparse{\pgfmathresult+0.5}}, mark repeat=15,mark phase=15,only marks,mark=triangle,mark size=1pt, red] table[x=Time,y=Single, col sep=comma,restrict x to domain=163:238]{data/mcp_adaptive_conflict.csv};

\draw [draw=black, fill=white] (axis cs:80,250) rectangle (axis cs:82,280);
\draw [draw=black, fill=white] (axis cs:160,250) rectangle (axis cs:162,280);

\addlegendentry{Single Address};
\addlegendentry{Conflicting Address};

\end{axis}
\end{tikzpicture}
  }
\end{figureA}

\cref{fig:mcp_all} shows the memory access time measured on an Intel Xeon D-1541 with different page policies.
The plot shows that closed-page policy can be distinguished from the other two using our method.
We also verified our results by reading out the \texttt{CLOSE\_PG} bit in the \texttt{mcmtr} configuration register of the integrated memory controller~\cite{IntelXeonE5v4Vol2}.

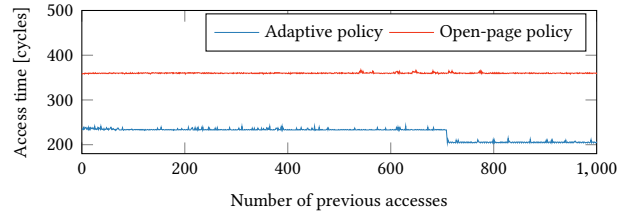
\begin{figureA}[t]{adaptive_policy}[Open-page policy and adaptive page policy can be distinguished by testing increasing numbers of accesses to the same row. The open-page policy (Intel Core i7-4790) always has the same timing for subsequent accesses, since the row always remains open.
The adaptive page policy (Intel Xeon E5-1630v4) only leaves the row open for a longer time after a larger number of accesses.]
  \resizebox {\hsize} {!} {
\begin{tikzpicture}
\begin{axis}[
mlineplot,
style={font=\footnotesize},
width={\hsize},
height=3.5cm,
xlabel={Number of previous accesses},
ylabel={Access time [cycles]},
xmin=0,
ymin=180,
ymax=500,
xmax=1000,
legend columns=-1,
]

\addplot+[no marks,mark size=1pt,PlotColorBlue] table[x=Pre,y=Time,col sep=comma]{data/adaptive_policy_lab08.csv};
\addlegendentry{Adaptive policy};

\addplot+[no marks,mark size=1pt,PlotColorRed] table[x=Pre,y=Time,col sep=comma]{data/adaptive_policy_lab02.csv};
\addlegendentry{Open-page policy};

\end{axis}
\end{tikzpicture}
}
\end{figureA}

We validated that we can distinguish open-page policy and adaptive page policy by running our experiments on two systems with the corresponding page policies.
\cref{fig:adaptive_policy} shows the results of these experiments.
The difference between open-page policy and adaptive policy is clearly visible.

Our experiments show that adaptive page policies often behave like closed-page policies.
This indicates the possibility of one-locating hammering on systems using an adaptive page policy.

\subsection{One-location Hammering on ARM}\label{sec:bridging_the_gap:arm}
To make \AttackName a more generic attack, it is essential to demonstrate it not only on Intel CPUs but also on ARM CPUs.
This is particularly interesting as ARM CPUs dominate the mobile market, and ARM-based devices are predominant also in IoT applications.
\citeA{Gruss2018Rowhammer} only demonstrated one-location hammering on Intel CPUs.
However, as one-location hammering is the most plausible hammering variant for \AttackName, we need to investigate whether it is possible to trigger one-location hammering bit flips on ARM.

In our experiments, we used a LG Nexus 4 E960 mobile phone equipped with a Qualcomm Snapdragon 600 (APQ8064)~\cite{APQ8064} SoC and 2GB of LPDDR2 RAM, susceptible to bit flips using double-sided hammering.
The page policy used by the memory controller is selected via the \texttt{DDR\_CMD\_EXEC\_OPT\_0} register: if the bit is set to 1, which is the recommended value~\cite{APQ8064EMCDS}, a closed-page policy is used.
If the bit is set to 0, an open-page policy is used.
Hence, we can expect the memory controller to preemptively close rows, enabling one-location hammering.

So far, bit flips on ARM-based devices have only been demonstrated in the combination of double-sided hammering, and uncached memory~\cite{Vanderveen2016} or access via the GPU~\cite{Frigo2018rowhammer}.
Even in the presence of a flush instruction~\cite{arm_arch_manualv8} or optimal cache eviction strategies~\cite{Lipp2016}, the access frequency to the two neighboring rows is too low to induce bit flips.
Furthermore, devices with the ARMv8 instruction set that allows exposing a flush instruction to unprivileged programs are usually equipped with LPDDR4 memory.

In our experiment, we allocated uncached memory using the Android ION memory allocator~\cite{Zeng2012}.
We hammered a single random address within the uncached memory region at a high frequency and then checked the memory for occurred bit flips.
We were able to observe \num{4} bit flips while hammering for \num{10} hours. 
Thus, we can conclude that there are ARM-based devices that are vulnerable to one-location hammering.

\subsection{Minimal Access Frequency for Rowhammer Attacks}\label{sec:evaluation:access-frequency}

A show stopper for \AttackName is if the frequency of memory accesses caused by processing network packets is not high enough to induce bit flips on one of our test systems successfully.
As the system performs many memory accesses when handling a network packet, the attacker, in fact, cannot tell whether only one location in a bank is hammered (\ie one-location hammering) or multiple locations (\ie single-sided hammering).
In particular, following the pigeon-hole principle, in our test setups with 32 bank (single DIMM) or 64 bank (dual DIMM) setups, we know that, if we access at least $n+1$ different addresses, \ie 33 or 65 respectively, at least two addresses must be served from the same bank.
Hence, we can assume that there is a good probability that the attacker actually does single-sided hammering.
Moreover, some addresses are accessed multiple times.

Previous work has investigated the minimal number of accesses that are necessary within a \SI{64}{\milli\second} refresh interval to still obtain bit flips.
\citeA{Kim2014} reported bit flips starting at \num{139000} row activations per refresh interval, which can be, depending on the page policy, identical to the number of memory accesses.
\citeA{Gruss2016Row} reported bit flips starting at \num{43000} and \citeA{Aweke2016} at \num{110000} memory accesses per refresh interval.

In our experiments, we send \SI{500}{\mega\bit/\second} (and more) over the network interface.
With a minimum size of \SI{64}{\byte} for Ethernet packets, we can send \SI{1024000}{packets} per second over a \SI{500}{\mega\bit/\second} connection.
As described in \cref{sec:evaluation:results}, we found functions which are called multiple times, \eg 6 times in the case of once function.
Hence, on a \SI{500}{\mega\bit/\second} connection, the attack can induce \SI{6144000}{accesses} per second.
Divided by the default refresh interval of \SI{64}{\milli\second}, we are at \SI{393216}{accesses} per refresh interval.
This is clearly above the previously reported required number of memory accesses~\cite{Gruss2016Row,Aweke2016,Kim2014}.
Hence, we conclude that in theory, if the system is susceptible to Rowhammer attacks, network packets can induce bit flips.
In the following section, we will describe how an attacker can exploit such bit flips.

\section{Exploiting Bit Flips over a Network}\label{sec:exploiting-bit-flips}

In this section, we discuss \AttackName attack scenarios to exploit bit flips over the network in detail.
We discuss the possible locations of bit flips in~\cref{sec:attack:flip-location}.
We describe different \AttackName attacks in~\cref{sec:attack:targets}.

\subsection{Bit Flip Location and Effect}\label{sec:attack:flip-location}
As the \AttackName attack does not control where in physical memory a bit flip is induced and, thus, what is stored at that location, the bit flip can lead to different consequences.
On a high level, we can divide bit flips into two groups, based on the location of the flip.
We distinguish between bit flips in user memory, \ie memory pages that are mapped as \texttt{user\_accessible} in at least one process, and bit flips in kernel memory, \ie memory pages that are never mapped as \texttt{user\_accessible}.
We can also distinguish the bit flips based on their high-level effect, again forming two groups.
The first group consists of bit flips that lead to a denial-of-service situation.
The second group consists of bit flips that do not lead to a denial-of-service situation.
If a denial-of-service situation is temporary, a system reboot may be necessary.
A denial-of-service situation can be persistent if the bit flip is written back to a permanent storage location.
Then it may be necessary to reinstall the system software or parts of it from scratch, clearly taking more time than just a reboot.
Denial-of-service attacks have a direct financial impact on companies due to unplanned downtimes and maintenance times.
Moreover, studies show that their announcement can also have a negative impact on the stock prices~\cite{Abhishta2017}.
Consequently, \AttackName poses a severe threat to servers vulnerable to the attack.

\subsection{Bit Flip Targets}\label{sec:attack:targets}
\AttackName may induce a bit flip in \textit{kernel memory}.
Depending on the modified location, parts of the operating system can behave unexpectedly, or the entire system may even halt.
Bit flips in \textit{user memory} may have similar consequences.

\subsubsection{File System Data Structures}
File system data structures, \eg inodes, are not directly part of the kernel code or data but are also in the kernel memory.
An inode is a data structure defining a file or a directory of a file system.
Each inode contains metadata such as the size of the file, owner and permission data as well as the disk block location of its data.
If a bit flips in the inode structure, it corrupts the file system and, thus, causes persistent loss of data.
This may again crash the entire system.

\subsubsection{SGX Enclave Page Cache}
If the victim machine supports Intel SGX~\cite{Costan2016}, an x86 instruction-set extension that allows the execution of programs in so-called \textit{secure enclaves} to run with integrity and confidentiality guarantees in untrusted environments, a bit flip easily causes a denial of service.
Enclave memory is stored in a physically contiguous block of memory that is encrypted using a Memory Encryption Engine~\cite{Gueron2016}.
\citeA{Jang2017SGXBomb} and \citeA{Gruss2018Rowhammer} demonstrated that if a bit flip in enclave memory is induced, the Memory Encryption Engine locks the memory controller, preventing any future memory operations and thus, halting the entire system.
While such a bit flip is not persistent itself, the unsafe halting of the entire system can leave permanent damage leading to a persistent denial-of-service.

\subsubsection{Application Memory in General}
If a bit flip occurs in memory of a user-space application, \eg code or data, a possible outcome is the crash of the program.
Such a flip may render the affected service unavailable.

Another outcome of a bit flip in the data of a user-space application, \eg in the database of a service, is that the service delivers modified, possibly invalid, content.
Depending on the service, its users cannot distinguish if the data is correct or has been altered.

\parhead{Altering DNS Entries to redirect to Malicious Services}\label{sec:attack:dns} 
To resolve domain names to the corresponding IP address, a DNS request~\cite{Mockapetris1987} is sent to a DNS server.
DNS servers are organized in a tree-like structure, building a distributed system to store DNS records.
A record consists of a type, a name, a class code, a time-to-live for caching, and the value.
For instance, the \texttt{A} record holds a 32-bit IPv4 address for a specific domain. 
However, DNS allows defining aliases to map one domain name to another.
This is used to define message transfer agents for a domain or to redirect domains.

In this attack, the attacker leverages \AttackName to induce a bit flip in a character of a DNS entry to make it point to a different domain.
For instance, \texttt{domain.com} changes to \texttt{dnmain.com} if the least-significant bit of the ``o'' character is flipped from `1' to `0'.
Such an attack is also referred to as bitsquatting~\cite{Dinaburg2011}.
Such bit flips in domains have been successfully exploited before using Rowhammer attacks~\cite{Razavi2016}. 
DNS zone transfers (AXFR queries) allow replicating DNS databases across different servers.
Using zone transfers, an attacker can retrieve entries of an entire zone at once.
The attacker queries the DNS server for its entries, mounts the attack and then verifies whether a bit flip at an exploitable position has occurred by monitoring changes in the queried entries.
If so, the attacker can register the changed domain and host a malicious service on the domain, \eg a fake website to steal login credentials or a mail server intercepting email traffic.
Users querying the DNS server for said entry connect to the server controlled by the attacker and are thus exposed to data theft.
A flip might also change an MX entry, pointing it to a different domain.
The attacker can then again register the domain and intercept connections that were intended to go to the original mail server.

\parhead{Rebuilding Trust in Revoked Certificates}\label{sec:attack:ocsp} 
An attacker can also target OCSP servers.
The Online Certificate Status Protocol (OCSP) is a protocol to retrieve the revocation status of a certificate~\cite{ocsprfc}.
In contrast to a certificate revocation list, where all revoked certificates are enumerated, the OCSP protocol enables to query the status of a single certificate.
This protocol shifts the workload from the user to the OCSP server, so that users, or more specifically browsers, do not have to store huge revocation lists.
Instead, the OCSP server manages a list of revoked certificate fingerprints.

Digital certificates are used to generate digital signatures that present the authenticity of digital documents or messages.
They are typically obtained from a trusted party, \eg a certificate authority.
The certificate allows verifying that a specific signature was indeed issued by the signer. 
However, if the corresponding private key of a certificate is exposed to the public, everyone can sign data in the name of the signer. 
Hence, a user can revoke a certificate to avoid any abuse.
\citeA{Liu2015PKI} evaluated \num{74} full IPv4 HTTPS scans and found that \num{8}\% of \num{38514130} unique SSL certificates served have been revoked.

To process a certificate validity request, the server queries its database for the requested certificate identifier. 
The result can either be that the certificate is revoked, not revoked (\ie valid), or that the state is unknown (\ie it is not in the database).
If a client tries to establish a secure connection to a server or check the validity of a signed document, it queries the OCSP server provided by the certificate. 
If the certificate has been revoked, the client aborts the connection or marks the signature as invalid.

In this attack, the attacker flips a bit in the memory of an OCSP server of a certificate authority where private keys of certificates have become public, and the certificates have thus been revoked. 
The attacker can either flip the status or the identifier of the certificate. 
As the status of the certificate is stored as an ASCII character in the OpenSSL OCSP responder~\cite{OpenSSLOCSPResponder}, one bit flip is sufficient to flip the ``R'' (revoked) to ``V'' (valid). 
Assuming the memory is filled with revocation list entries, which are on average \SI{100}{B} for this specific responder, an attacker has a chance of \SI{0.125}{\percent} \emph{per bit flip} to make a random certificate valid again. 
Thus, an attacker can again reuse that certificate (with the known private key) to sign documents or data and, thus, impersonate the original signer. 

A weaker, but more likely attack scenario, is to flip a bit in the certificate identifier. 
Such a bit flip leads to the OCSP server not finding the certificate in its database anymore, thus, returning ``unknown'' as the state. 
Most browsers fall back to their own certificate revocation list in such a case~\cite{Mutton2014OCSP,Langley2014Revocation,Goodwin2015OCSP}. 
However, only high-value revocations are kept in the browser's list, making it very unlikely that the certificate is in the certificate revocation list of the browser~\cite{Langley2014Revocation}.
Hence, an attacker can again reuse that certificate. 

\parhead{Other attacks}
The attack scenarios described above are by far not exhaustive. 
With bit flips in applications, attackers have numerous possibilities to modify random data, yielding different, disastrous consequences.
However, the outlined attacks highlight the severity of remotely induced bit flips by \AttackName.

\subsubsection{Cryptographic Material}
Cryptographic material as part of the application memory is particularly interesting for attacks.
In the past, it has been demonstrated that fault attacks on RSA public keys result in broken keys which are susceptible to key factorization~\cite{Brier2006, Berzati2008}. 
Therefore, also public key material has to be protected against faults.
\citeA{Muir2006} remarked that a bit flip in an RSA public key allows an attacker with a non-negligible probability to compute a private key corresponding to the modified key in a reasonable amount of time.
Thus, an attacker can flip a bit of a public RSA key in memory using \AttackName, giving the attacker the same privileges and permissions as the owner of the original key.
These permissions are only temporary, \eg until the key is reloaded from the hard drive. 

\parhead{Distribution of Malicious Software on Version-Control Hosting Services}\label{sec:attack:public-key}
An attacker can compromise a hosting service to distribute malicious software. 
The number of organizations using hosting services for revision control to manage changes to their source code, documents or other information is increasing steadily.
These services can either be subscription based, \eg GitHub~\cite{github}, or self-hosted, \eg GitLab~\cite{gitlab}, and, thus, are deployed on many web servers to distribute their software.

To commit changes to a version-controlled repository, users authenticate with the service using public-key cryptography.
Typically, users generate an SSH key pair~\cite{Ylonen2006}, \eg using RSA~\cite{RSA}, upload the public key to the service, and store the private key securely on their local system.
As the position of the bit flip cannot be controlled using \AttackName, an attacker can improve the probability to induce a bit flip in the modulus of a public key by loading as many keys as possible into the main memory of the server.
Some APIs, \eg the GitLab API~\cite{gitlabAPI}, allow enumerating the users registered for the service as well as their public keys.
By enumerating and, therefore, accessing all public keys of the service, the attacker loads the public keys into the DRAM.

In the first step of the attack, the attacker enumerates all keys of all users and stores them locally.
In the second step, the attacker mounts \AttackName to induce bit flips on the targeted system. 
The more keys the attacker loaded into memory, the more likely it is that the bit flip corrupts the modulus of a public key of a user.
For instance, with \SI{80}{\percent} of the memory filled with \num{4096}-bit keys, the chance to hit a bit of a modulus is \SI{79.7}{\percent}.
As the attacker does not know which key was affected by the bit flip, the attacker enumerates all keys again and compares them with the locally stored keys.
If a modified key has been found, the attacker computes a new corresponding private key~\cite{Muir2006,Razavi2016}.
The attacker uses this new key to authenticate with the service, impersonating the user. 
Consequently, the attacker can make changes to the software repository as that user and, thus, introduce bugs that can later be exploited if the software is distributed.
The original public key will be restored after a while when the key is evicted from the page cache and has to be reloaded from the hard drive.
As the correct key is restored, the attack leaves no traces. 
Furthermore, it also breaks the non-repudiation guarantee provided by the public-key authentication, making the victim whose public key was attacked the prime suspect in possible investigations.

\section{Evaluation}\label{sec:evaluation}

In this section, we evaluate \AttackName and its performance.
We show that the number of bit flips induced by \AttackName depends on how the cache is bypassed and the memory-controller's page policy.
We evaluate which kernel functions are executed when handling a UDP network packet.
We describe the bit flips we obtained when running \AttackName in different attack scenarios.
Finally, we show that TRR does not protect against \AttackName or Rowhammer in general.

\subsection{Environment}\label{sec:evaluation:environment}

\begin{table*}[t]
  \caption{List of test systems that were used for the experiments.}
  \label{tab:list-of-devices}
  \begin{tabular}{cllll}
    \toprule
    \thead{Device} & \thead{CPU} & \thead{DRAM} & \thead{Network card} & \thead{Operating system}\\
    \midrule
    Desktop & Intel i7-6700K @ \SI{4}{\giga\hertz} & \SI{8}{\giga\byte} DDR4 @ \SI{2133}{\mega\hertz} & Intel 10G X550T & Ubuntu 16.04\\
    \hdashline
    Server & Intel Xeon E5-1630v4 @ \SI{3.7}{\giga\hertz}  & \SI{8}{\giga\byte} DDR4 @ \SI{2133}{\mega\hertz} & Intel i210/i218-LM Gigabit & Xubuntu 17.10\\
    \hdashline
    Server & Intel Xeon D-1541 @ \SI{2.1}{\giga\hertz} & \SI{8}{\giga\byte} DDR4 @ \SI{2133}{\mega\hertz}  & Intel i350-AM2 Gigabit & Ubuntu 16.04 \\
    \hdashline
    LG Nexus 4 & Qualcomm APQ8064 @ \SI{1.5}{\giga\hertz} & \SI{2}{\giga\byte} LPDDR2 @ \SI{533}{\mega\hertz}  & USB Adapter & Android 5.1.1 \\
    \bottomrule
  \end{tabular}
\end{table*}

In our evaluation, we used the test systems listed in~\cref{tab:list-of-devices}. 
We used the first system for our experiments with a non-default network driver implementation that uses \texttt{clflush} in the process of handling a network packet, and the second and third system for our experiments with Intel CAT. 
To mount \AttackName, we used a Gigabit switch to connect two other machines with the victim machine. 
The two other machines were used to flood the victim machine with network packets triggering the Rowhammer bug. 
We used the fourth system for our experiments on an ARM-based device that uses uncached memory in the process of handling a network packet.

\subsection{Evaluation of the Different Cache Bypasses for \AttackName}\label{sec:evaluation:results}

In~\cref{sec:bridging_the_gap}, we investigated the requirements to trigger the Rowhammer bug over the network.
In this section, we evaluate \AttackName for the three cache-bypass techniques (see~\cref{sec:attack:overview}):
a kernel driver that flushes (and reloads) an address whenever a packet is received, Intel Xeon CPUs with Intel CAT for fast cache eviction, and uncached memory on an ARM-based mobile device.

\parhead{Driver with \texttt{clflush}}
To verify that \AttackName can induce bit flips, we used a non-default network driver implementation that uses \texttt{clflush} in the process of handling a network packet on an Intel i7-6700K CPU.
We sent UDP packets with up to \SI{500}{\mega\bit\per\second} and scanned memory regions where we expected bit flips.
We observed a bit flip every \SI{350}{\milli\second} showing that hammering over the network is feasible if at least two memory accesses are served from main memory, due to flushing an address while handling a network packet.
Thus, in this scenario, up to \ValBitflipRateMax~bit flips per hour can be induced.

\parhead{Eviction with Intel CAT}
The operating system will handle every network packet received by the network card.
The operating system parses the packets depending on their type, validates their checksum and copies and delivers every packet to each registered socket queue.
Thus, for each received packet quite some code is executed before the packet finally arrives at the application destined to handle its content.

We tested \AttackName on Intel Xeon CPUs with Intel CAT.
The number of cache ways has been limited to a single one for code handling the processing of UDP packets, resulting in fast cache eviction.
If a function is called multiple times for one packet, the same addresses are accessed and loaded from DRAM with a high probability, thus, hammering this location.
To estimate how many different functions are called and how often they are called, we use the \textit{perf} framework~\cite{linux_perf} to count the number of function calls related to UDP packet handling.
\cref{sec:appendix:kernel-addresses} shows the results of a system handling UDP packets.
Out of \num{27} different functions we identified, most were called only once for each received packet.
The function \texttt{\_\_udp4\_lib\_lookup} is called twice.
In a more extensive profiling scan, we found that \texttt{nf\_hook\_slow} is called 6 times while handling UDP packets on some kernels.

With this knowledge, we analyzed how many bit flips can be induced from this code execution.
We observed \num{45} bit flips per hour on the Intel Xeon E5-1630v4.
As TRR is active on this system (see~\cref{sec:evaluation:trr}), fewer bit flips occur in comparison to systems without TRR.
In~\cref{sec:evaluation:page-policy-influence}, we evaluate the number of bit flips on the Intel Xeon D-1541 depending on the configured page policy.

\parhead{Uncached Memory}
In~\cref{sec:bridging_the_gap:arm}, we demonstrated that ARM-based devices are vulnerable to one-location hammering in general.
To investigate whether bit flips can also be induced over the network, we connect the LG Nexus 4 using an OTG USB ethernet adapter to a local network.
Using a different machine, we send as many network packets as possible to the mobile phone.
An application on the phone allocates memory and repeatedly checks the allocated memory for occurred bit flips.
However, we were not able to observe any bit flips on the device within 12 hours of hammering.
As the device does not deploy a technology like Intel CAT (\cref{sec:background:caches:cat}), the cache is not limited for certain applications and, thus, the eviction of code or data used by handling memory packets has a low probability.
As network drivers often use DMA memory and, thus, uncached memory, bit flips induced by the network are more likely if the network driver itself uses uncached memory.
While we identified a remarkable number of around \num{5500} uncacheable pages used by the system, we were not able to induce any bit flips over the network.
However, we found that the USB ethernet adapter only allowed for a network capacity of less than \SI{16}{\mega\bit\per\second}, which is clearly too slow for a \AttackName attack.
It is very likely that with a faster network connection, \eg more than \SI{200}{\mega\bit\per\second}, it is possible to induce bit flips.
Nevertheless, we were successfully able to induce bit flips using \AttackName on the Intel Xeon E5-1630v4 where one uncached address is accessed for every received UDP packet.

\subsection{Influence of Memory-Controller Page Policies on Rowhammer}\label{sec:evaluation:page-policy-influence}

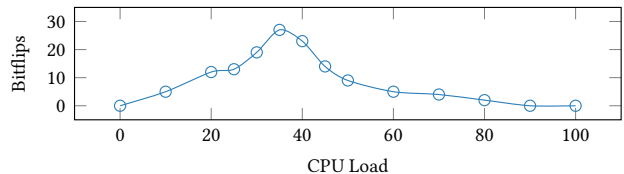
\begin{figureA}[t]{bitflip_cpu_load}[Number of bit flips depending on the CPU load with a closed-page policy after 15 minutes (Xeon D-1541).]
  \resizebox {\hsize} {!} {
	\begin{tikzpicture}
\begin{axis}[
mlineplot,
style={font=\footnotesize},
width={\hsize},
height=3cm,
xlabel={CPU Load},
ylabel={Bitflips},
ymin=-5,
ymax=35,
]
\addplot+[mark=o,smooth,PlotColorBlue] table[x=Load,y=Bitflips, col sep=comma]
  {data/bitflip_cpu_load.csv};

\end{axis}
\end{tikzpicture}

  }
\end{figureA}

In order to evaluate the actual influence of the used memory-controller page policy on \AttackName, \ie how many bit flips can be induced depending on the policy used, we mounted the \AttackName in different settings.
The experiment was conducted on our Intel Xeon D-1541 test system, as the BIOS of its motherboard allowed to chose between different page policies: \textit{Auto}, \textit{Closed}, \textit{Open}, \textit{Adaptive}.
For each run, we configured the victim machine with one of the policies and Intel CAT, and, mounted a \AttackName attack for at least \num{4} hours.
To detect bit flips, we ran a program on the victim machine that mapped a file into memory.
The program then repeatedly scans the content of all allocated pages and reports bit flips if the content has changed.

We detected 11 bit flips in 4 hours with the \textit{Closed} policy, with the first one after \num{90} minutes.
We did not observe any bit flips with the \textit{Open} policy within the first \num{4} hours.
However, when running the experiment longer, we observed \num{46} bit flips within 10 hours.
With the \textit{Adaptive} policy, we observed \num{10} bit flips in 4 hours, with the first one within the second hour of the experiment.
While this experiment was conducted without any additional load on the system, we see in~\cref{fig:bitflip_cpu_load} that additional CPU utilization increases the number of bitflips drastically.
Using the \textit{Closed} policy, we observed 27 bitflips with a load of 35\% within 15 minutes.

These results do not immediately align with the assumption that a policy that preemptively closes rows is required to induce bit flips using one-location hammering.
However, depending on the addresses that are accessed and the constant eviction through Intel CAT, it is possible that two addresses map to the same bank but different rows and, thus, bit flips can be induced through single-sided hammering.
In fact, the attacker cannot know whether the hammering was actually one-location hammering or single-sided hammering.
However, as long as a bit flip occurs, the attacker does not care how many addresses mapped to the same bank.
Finally, depending on the actual parameters used by a fixed-open-page policy, a row can still be closed early enough to induce bit flips.

\subsection{Bit Flips induced by \AttackName}\label{sec:evaluation:bit-flips}

As described in~\cref{sec:attack:flip-location}, a bit flip can occur in user space or kernel space leading to different effects depending on the memory it corrupts.
In this section, we present bit flips that we have observed in our experiments and the effects they have caused.

\parhead{Kernel image corruption and kernel crashes}
We observed \AttackName bit flips that caused the system not to boot anymore.
It stopped responding after the bootloader stage.
We inspected the kernel image and compared it to the original kernel image distributed by the operating system.
As the kernel image differed blockwise at many locations, we assume that the \AttackName caused a bit flip in an inode of the file system.
The inode of a program that wanted to write data did not point to the correct file any longer but to the kernel image and, thus, corrupted the kernel image.

Furthermore, we observed several bit flips immediately halting the entire system such that interaction with it was not possible any longer.
By debugging the operating system over a serial connection, we detected bit flips in certain modules such as the keyboard or network driver.
In these cases, the system was still running but did not respond to any user input or network packets anymore.
We also observed bit flips that were likely in the SGX EPC region, causing an immediate permanent locking of the memory controller.

\parhead{Bit flips in user space libraries and executables}
We observed that bit flips crashed running processes and services or prevented the execution of others as the bit flip triggered a segmentation fault when functions of a library were executed.
On one occasion, a bit flip occurred either in the SSH daemon or the stored passwords of the machine, preventing any user to login on the system.
The system was restored to a stable state only by rebooting the machine and thus reloading the entire code from disk.

We also validated that an attacker can increase chances to flip a bit in a target page by increasing the memory usage of a user program.
In fact, this was the most common scenario, overlapping with our general test setup to detect bit flips for our evaluation.
Unsurprisingly, these bit flips equally occur when filling the memory with actual contents that the attacker targets.

\subsection{Target Row Refresh (TRR)}\label{sec:evaluation:trr}

Previous assumptions on the Rowhammer bug lead to the conclusion that only bit flips in the victim row adjacent to the hammering rows would occur.
While the probability for bit flips to occur in directly adjacent rows is much higher, \citeA{Kim2014} already showed rows further away (even a distance of 8 rows and more) are affected as well.
Still, the hardware vendors opted for implementing countermeasures focusing on the directly adjacent rows.

With the Low Power Double Data Rate 4 (LPDDR4) standard, the \citeA{LPDDR4} defines a reliability feature called Target Row Refresh (TRR).
The idea of TRR is to refresh adjacent rows if the targeted row is accessed at a high frequency.
More specifically, TRR works with a maximum number of activations allowed during one refresh cycle, the maximum active count.
Thus, if a double-sided Rowhammer attack (\cref{sec:background:rowhammer}) is mounted, and two hammered rows are accessed more than the defined maximum active count, the adjacent rows (in particular the victim row of the attack) will be refreshed.
As the potential victim rows are refreshed, in theory, no bit flip will occur, and the attack is mitigated.
However, in practice, bit flips can be further away from the hammered rows and thus TRR may be ineffective.

With the Ivy Bridge processor family, Intel introduced Pseudo Target Row Refresh (pTRR) for Intel Xeon CPUs to mitigate the Rowhammer bug~\cite{Kaczmarski2014}.
On these systems pTRR-compliant DIMMs must be used; otherwise, the system will default into double refresh mode, where the time interval in which a row is refreshed is halved~\cite{Kaczmarski2014}.
However, \citeA{Kim2014} showed that a reduced refresh period of \SI{32}{\milli\second} is not sufficient enough to impede bit flips in all cases.
While pTRR is implemented in the memory controller~\cite{Mandava2017}, DRAM module specifications theoretically allow automatically running TRR in the background~\cite{MicronDDR4}.

In our experiments, we were able to induce bit flips on a pTRR-supporting DDR4 module using double-sided hammering on an Intel i7-6700K.
The bit flips occurred in directly adjacent rows and rows further away.
We observed that when using the same DDR4 DRAM on the Intel Xeon E5-1630 v4 CPU, no bit flips occurred in the directly adjacent rows, but we observed no statistically significant difference in the number of bit flips for the rows further away.
This indicates that TRR is active on the second machine but also that TRR does not prevent the occurrence of exploitable bit flips in practice.
Thus, we conclude that the TRR hardware countermeasure is insufficient in mitigating Rowhammer attacks.

\section{Countermeasures}\label{sec:countermeasures}

Since \AttackName does not require any attack code in contrast to a regular Rowhammer attack, \eg no attacker-controlled code on the system, most countermeasures do not prevent our attack.

Countermeasures based on static code analysis aim to detect attack code in binaries~\cite{Irazoqui2016mascat}.
However, as our attack does not use any suspicious code and does not execute a program, these countermeasures do not detect the ongoing attack.
Other countermeasures detect on-going attacks using hardware- and software-based performance counters~\cite{Herath2015, Payer2016, Gruss2016Flush, Corbet2016, Chiappetta2015,Zhang2016CloudRadar} and subsequently stop the corresponding programs.
However, when hammering over the network, the large amount of memory accesses are executed by the kernel itself, and the kernel cannot just be terminated or stopped like a regular program.
Hence, these countermeasures cannot cope with our attack.
Modifying the system memory allocator to hinder the exploitability of bit flips~\cite{Vanderveen2016,Gruss2016Row,Brasser2017catt} may generally work against \AttackName.
However, the hammering is in practice done by the kernel, so the proposed isolation schemes are ineffective, and new schemes have to be proposed.

ANVIL~\cite{Aweke2016} uses performance counters to detect and subsequently mitigate Rowhammer attacks.
Since ANVIL, in its current form, does not detect one-location hammering~\cite{Gruss2018Rowhammer}, it also does not detect our attack.
While we believe an adapted version of ANVIL could detect our attack, it would require evaluating whether the false positive and false negative rates allow for an application in practice.
B-CATT~\cite{Brasser2017catt} blacklists vulnerable locations, thus, effectively reducing the amount of usable memory, but fully eliminating the Rowhammer bug.
B-CATT would work against \AttackName, but previous work has found that it is not practical as it would block too much memory~\cite{Kim2014,Gruss2018Rowhammer}.

In general, we recommend reviewing any network stack and network services code.
Uncached memory and \texttt{clflush} instructions should only be used with extreme care, and it may even be necessary to add artificial slow downs such that they cannot be exploit for \AttackName attacks anymore.
If this is not possible for technical reasons, the threat model of the device should be revisited and reevaluated. 
Mitigating our eviction-based \AttackName attack might be more straight-forward, as it requires a specific configuration for Intel CAT.
Either avoiding the restriction to a low number of cache ways via Intel CAT on network-connected systems or installing ECC memory would likely be sufficient to make our attack very improbable to succeed.
Hence, we also recommend using Intel CAT very carefully in network-connected systems.

\section{Discussion}\label{sec:discussion}

\parhead{Hardware requirements}
To induce the Rowhammer bug, one needs to access memory in the main memory repeatedly and, thus, needs to circumvent the cache.
Therefore, either native flush instructions~\cite{Yarom2014}, eviction~\cite{Gruss2016Row,Aga2017} or uncached memory~\cite{Vanderveen2016} can be used to remove data from the cache.
In particular, for eviction-based \AttackName, the system must use Intel CAT as described in~\cref{sec:background:caches:cat} in a configuration that restricts the number of ways available to a virtual machine in a cloud scenario to guarantee performance to other co-located machines~\cite{CATPerformance2015}.
If none of these capabilities are available over the network, an attacker could not mount \AttackName in practice.

One limitation of our attack is that only DRAM susceptible to bit flips can be exploited using a Rowhammer attack and, thus, \AttackName.
To reduce the risk of bit flips on servers, one would assume that cloud providers tend to deploy ECC RAM usually.
However, many cloud providers offer to rent hardware without ECC RAM~\cite{HetznerDedicatedRoot,DefineQualityDedicatedRoot,MyLocDedicatedRoot,webtropicDedicatedRoot,netcupDedicatedRoot,fasthostsDedicatedRoot}, potentially allowing \AttackName attacks.
DRAM with ECC can only be used in combination with Intel Xeon CPUs and can detect and correct 1-bit errors.
Therefore it can deal with single bit flips.
While non-correctable multi-bit flips can be exploitable~\cite{Aichinger2015hpec,Aichinger2015memcon,Lanteigne2016}, they often end up in a denial-of-service attack depending on the operating system's response to the error.

\parhead{Network traffic}
\AttackName sends as many network packets to the victim machine as possible, aiming to induce bit flips.
Depending on the actual attack scenario (see~\cref{sec:exploiting-bit-flips}), additional traffic, \eg by enumerating the public keys of the service, is generated.
If the victim uses network monitoring software, the attack might be detected and stopped, due to the highly increased amount of traffic.
In our experiments, we sent a stream of UDP packets with up to \SI{500}{\mega\bit/\second} to the target system.
We were able to induce a bit flip every \SI{350}{\milli\second} and, thus, if the first random bit flip already hits the target or causes a denial-of-service, the attack could already be successful.
However, as the rows are periodically refreshed, an attacker only needs an extraordinary high burst of memory accesses to a row between two refreshes, \ie within a period of \SI{64}{\milli\second}.
Hence, an attacker could mount \AttackName for a few hundred milliseconds and then pause the attack for a longer time.
These short network spikes may circumvent network monitoring software that might otherwise detect and prevent the on-going attack, \eg by null routing the victim server.

\parhead{Gigabit LTE on Mobile Devices}\label{sec:discussion:gigabit-lte} 

While ethernet adapters in mobile phones are uncommon, many ARM-based embedded devices in IoT setups are equipped and connected with gigabit ethernet.
However, we expect the maximum throughput of these network cards to be too low on many of these devices, \eg the Raspberry Pi 3 Model B+~\cite{raspberrypi3bp}, and also WiFi chips typically offer too little capacity.
However, on more recent processors, \eg the Qualcomm Snapdragon 845 chipset~\cite{snapdragon845}, and modems like the Qualcomm X20 Gigabit LTE modem, throughputs up to \SI{1.2}{\giga\bit\per\second} are possible in practice.
This would enable to send enough packets to hammer specific addresses to induce bit flips on the device and, thus, to successfully mount \AttackName.

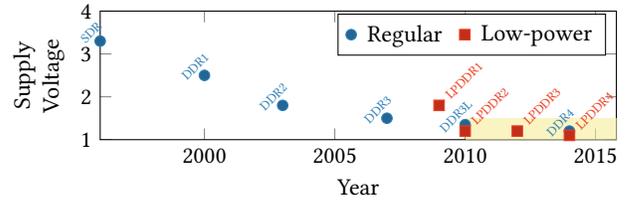
\begin{figureA}[t]{voltage_years}[Minimum DRAM supply voltages for different DDR standards. The highlighted area marks the voltage and manufacturing years of DRAM modules where Rowhammer bit flips have been reported.]
  \resizebox {\hsize} {!} { \begin{tikzpicture}
\pgfplotsset{every node near coord/.append style={font=\tiny,rotate=45}}
\begin{axis}[legend columns=2, xlabel=Year, ylabel=Supply Voltage, ylabel style={text width=1.5cm,align=center},width=\hsize, height=3.3cm, xtick={2000,2005,2010,2015}, xticklabel style={/pgf/number format/1000 sep=},ymax=4,xmin=1996,ymin=1]
\addplot+[only marks,nodes near coords,point meta=explicit symbolic] table[x=Year,y=Minimum,meta=Standard] {data/voltage_years_ddr.csv};
\addlegendentry{~Regular~\quad~\quad}
\addplot+[only marks,nodes near coords,point meta=explicit symbolic,every node near coord/.style={font=\tiny,rotate=45,yshift=-1.25ex,xshift=1.5em}] table[x=Year,y=Minimum,meta=Standard] {data/voltage_years_lpddr.csv};
\addlegendentry{~Low-power}
\draw[fill=yellow,opacity=0.6,draw=none] (axis cs:2010,0) rectangle (axis cs:2020,1.5);
\end{axis}
\end{tikzpicture}
  }
\end{figureA}

\parhead{Influence of DRAM Supply Voltage on Rowhammer Effect} 
%
Kim~\etal\cite{Kim2014} identified voltage fluctuations as the root cause of DRAM disturbance errors, \eg the Rowhammer bug.
However, no study so far has investigated the direct effect of the DRAM supply voltage on Rowhammer bit flips.
In fact, we can already observe a direct correlation between a low DRAM supply voltage by reviewing related work.
\cref{fig:voltage_years} shows how the DRAM voltage has been reduced over the past years.
Previous work observed that the vulnerability of DRAM modules is related to the manufacturing date, \ie no bitflips before 2010~\cite{Kim2014,Seaborn2015BH}.
However, as shown in \cref{fig:voltage_years} there are at least two possible correlations with the Rowhammer bug, the manufacturing date, and the supply voltage.

Indeed, Rowhammer has only been reported on DRAM modules with a voltage below 1.5 volts~\cite{Kim2014,Seaborn2015BH}, \ie DDR3~\cite{Kim2014,Seaborn2015BH}, DDR4~\cite{Pessl2016}, LPDDR2 and LPDDR3~\cite{Vanderveen2016}, and LPDDR4~\cite{VanDerVeen2016CCSPresentation}.

We investigated the influence of the DRAM voltage on the occurrence of bit flips on two systems.
We tested voltage increases in \SI{0.01}{V} steps.
On three tested systems ($2\times$ DDR4, $1\times$ DDR3), we observed no significant change in the number of bit flips, \ie the number of bit flips stayed in the same order of magnitude, even when increasing the voltage by \SI{0.2}{V}.
Future work should investigate whether other voltage-related parameters could lead to a straightforward elimination of the Rowhammer bug.

\section{Conclusion}\label{sec:conclusion}
In this paper, we presented \AttackName, the first truly remote Rowhammer attack, without a single attacker-controlled line of code on the targeted system.
We demonstrate attacks on systems that use uncached memory or flush instructions while handling network requests, and systems that don't use either but are protected by Intel CAT.
In all cases, we were able to induce multiple bit flips per hour on real-world systems, leading to temporary or persistent damage on the system.
We showed that depending on the location, the bit flip compromises either the security and integrity of the system and the data of its users.
In some cases, the system was rendered unbootable after the attack.

We presented a method to automatically identify the page policy used by the memory controller.
Consequently, we found that adaptive page policies are also vulnerable to one-location hammering.
While we were able to mount the first one-location hammering attack on an ARM device, the network capacity on this device was too low for \AttackName.

Transforming formerly local attacks into remote attacks is always a landslide in security.
Assumptions that were true for the local scenario are largely invalid in a remote scenario.
In particular, all defenses and mitigation strategies were designed against local Rowhammer attacks, \ie remote Rowhammer attacks were out of scope.
Hence, \AttackName requires the re-evaluation of the security of millions of devices where the attacker is not able to execute attacker-controlled code.
Finally, our work demonstrates that we need to develop countermeasures with the root cause of both local and remote Rowhammer attacks in mind.

\section*{Acknowledgement}\label{sec:acknowledgement}
We thank Mattis Turin-Zelenko and Paul Höfler for help with some experiments.
We thank Stefan Mangard, Anders Fogh, Thomas Dullien, and Jann Horn for fruitful discussions.

This work has been supported by the Austrian Research Promotion Agency (FFG) via the K-project DeSSnet, which is funded in the context of COMET – Competence Centers for Excellent Technologies by BMVIT, BMWFW, Styria and Carinthia.
This project has received funding from the European Research Council (ERC) under the European Union’s Horizon 2020 research and innovation programme (grant agreement No 681402).

%
\bibliographystyle{ACM-Reference-Format}
\bibliography{main}  
%
%
\appendix

\section{Kernel Accesses for Network Packets}
\label{sec:appendix:kernel-addresses}

\cref{tab:funccount} shows the results of the \textit{funccount} script of the \textit{perf} framework~\cite{linux_perf} for functions with udp in their name while the targeted system is flooded with UDP packets.

\begin{table}[t]
  \caption{Results of \textit{funccount} on the victim machine for functions with udp in their name while the system is flooded with UDP packets.}
  \label{tab:funccount}
  \begin{tabular}{lr}
    \toprule
    \thead{Function} & \thead{Number of calls}\\
    \midrule
    \texttt{\_\_udp4\_lib\_lookup}  & \num{2000024} \\
    \texttt{\_\_udp4\_lib\_rcv}     & \num{1000012} \\
    \texttt{udp4\_gro\_receive}     & \num{1000012} \\
    \texttt{udp4\_lib\_lookup\_skb} & \num{1000012} \\
    \texttt{udp\_error}             & \num{1000012} \\
    \texttt{udp\_get\_timeouts}     & \num{1000013} \\
    \texttt{udp\_gro\_receive}      & \num{1000013} \\
    \texttt{udp\_packet}            & \num{1000012} \\
    \texttt{udp\_pkt\_to\_tuple}    & \num{1000012} \\
    \texttt{udp\_rcv}               & \num{1000012} \\
    \texttt{udp\_v4\_early\_demux}  & \num{1000012} \\
    \bottomrule
  \end{tabular}
\end{table}

\end{document}